\documentstyle[aas2pp4]{article}

\input{psfig}

\lefthead{Brandner et al.}
\righthead{}

\def\NII{N{\sc ii}}

\begin{document}

\title{The Hourglass Nebulae of Sher\,25 and SN\,1987\,A: Two of a Kind?\footnote{Based on 
observations obtained at CTIO, at the European Southern Observatory, La Silla
(ESO Proposal No.\ 58.E-0965, 59.D-0330), and on observations made with the 
NASA/ESA Hubble Space Telescope, obtained from the data archive at the Space 
Telescope Science Institute. STScI is operated by the Association of 
Universities for Research in Astronomy, Inc., under the NASA contract NAS
       5-26555. }}

\author{Wolfgang Brandner\altaffilmark{2,3}, You-Hua Chu\altaffilmark{2,4},
Frank Eisenhauer\altaffilmark{5}, Eva K.\ Grebel\altaffilmark{3}, Sean D.\ Points\altaffilmark{2}}

\affil{$^2$University of Illinois at Urbana-Champaign, Department of Astronomy,
1002 West Green Street, Urbana, IL 61801, USA}
\authoremail{brandner@astro.uiuc.edu, chu@astro.uiuc.edu, points@astro.uiuc.edu}

\affil{$^3$Astronomisches Institut der Universit\"at W\"urzburg, Am Hubland,
D-97074 W\"urzburg, Germany}
\authoremail{grebel@astro.uni-wuerzburg.de}

\affil{$^4$Visiting astronomer, Cerro Tololo Inter-American Observatory,
NOAO, which are 
operated by AURA, Inc.,
under contract with the NSF}

\affil{$^5$Max-Planck-Institut f\"ur extraterrestrische Physik, Giessenbachstra{\ss}e, D-85740 Garching, Germany}
\authoremail{eisenhau@mpa-garching.mpg.de}

\begin{abstract} 
We have performed a detailed study of the morphology and kinematics
of the hourglass-shaped nebula around the blue supergiant Sher 25
in the galactic giant H{\sc ii} region NGC 3603. Near-infrared high
resolution adaptive optics images in the Br$\gamma$ line and HST/NICMOS
images in the He{\sc i}\,1.08$\mu$m line were compared with iso-velocity
maps in the H$\alpha$ and [N{\sc ii}] lines.
 
The adaptive optics observations clearly resolved the width of the ring
(0\farcs9, i.e., 0.027 pc), yielding $\delta$R/R=1:8. We show that the
H$\alpha$ and [N{\sc ii}] lines trace the entire silhouette of the
hourglass. The bipolar lobes of the hourglass expand at 70 km\,s$^{-1}$, 
whereas the ring around the waist of the hourglass expands at 30 
km\,s$^{-1}$. Both the ring and the bipolar lobes have about the same
dynamical age, indicating a common origin and a major outburst 
and mass-loss event 6630 yr ago. The ionized mass 
within the hourglass is between 0.3 M$_\odot$ and 0.6 M$_\odot$ - 
quite comparable to the total mass suggested for the 
expanding (pre-supernova) shell around SN\,1987\,A.
 
The hourglass structure around Sher 25 is similar to that of 
SN\,1987\,A in spatial extent, mass, and velocities. The major
differences between these two nebulae might arise from environmental
effects.  Both internal and external ionization sources are available
for Sher 25's nebula. Furthermore, Sher 25 and its hourglass-shaped
nebula appear to be moving to the south-west with respect to the ambient 
interstellar
medium, and ram pressure has apparently deformed the hourglass. We
conclude that the circumstellar nebulae around SN\,1987\,A and Sher 25
are very similar and define a new class of nebulae
around blue supergiants in their final evolutionary stage.

\end{abstract}


\section{Introduction}

The H{\sc ii} ring around Sher 25  was serendipitously discovered
(Brandner et al.\ 1997) in the course of a high-spectral resolution study 
of emission line knots in the giant H{\sc ii} NGC 3603 region (distance 
6 kpc, e.g., Clayton 1986). 
It has a diameter of 0.4 pc, is tilted by 65$^\circ$ against
the plane of the sky, and is expanding. Perpendicular to the plane of the ring 
are two extended nebular structures which appear to be bipolar lobes.
The high [\NII]/H$\alpha$ ratio of 2.1:1 in the nebular material 
gives evidence that we see circumstellar material enriched by the CNO cycle and
ejected by Sher 25 (Brandner et al.\ 1997).

The present study aims at investigating the physical structure
and dynamics of the ring and bipolar outflow around Sher 25. Through a
comparison to SN\,1987\,A's rings we hope to gain a better
understanding of the basic physics and formation mechanisms of bipolar 
nebulae around blue supergiants.

\section{Observations and data reduction} 

Low-dispersion spectra of Sher 25's nebula 
were obtained on 1997 February 3 with the
ESO/MPI 2.2m telescope and EFOSC2. The slit width was 1\farcs5.
The spectra have a resolution of 0.2 nm pixel$^{-1}$ and cover the wavelength 
range from 517 nm to 928 nm. They were wavelength and flux calibrated 
using IRAF\footnote{IRAF is distributed by the National Optical Astronomy 
Observatories (NOAO)}.

Long-slit echelle spectroscopic mapping of Sher 25's nebula 
was carried out on 1997 February 27 at the CTIO 4m telescope. The 
cross disperser of the echelle spectrograph had been replaced by a flat mirror.
To map Sher 25's nebula, we used an east-west oriented slit (slit width 
1\farcs65) and observed 14 slit positions, spaced by 2$''$ from
12$''$ north to 14$''$ south of Sher 25.
The integration time per slit position was 10 min and the
velocity resolution $\approx$ 3.8 km\,s$^{-1}$ pixel$^{-1}$. 
The final spatial resolution in east-west direction, i.e.\
along the slit was 1\farcs7 (seeing), and 4$''$ in north-south
direction.
Using IRAF we flat-fielded, rectified, and wavelength calibrated the 2-D frames.
They then were stacked into a 3-D data cube and analyzed with IDL.

Adaptive optics imaging of Sher 25 and its nebula
was carried out with ADONIS/SHARP at the ESO 3.6m telescope
on 1997 May 19. We observed with a Fabry-Perot-Etalon 
($\lambda/\delta\lambda$=2200)
centered on Br$\gamma$ (2.166$\mu$m) and the nearby continuum
(2.157$\mu$m and 2.174$\mu$m). For both line and continuum we obtained 
$40\times 1$\,min exposures. Due to a very short coherence time of
the atmospheric turbulence, the spatial resolution achieved was not 
diffraction limited but 0\farcs35. The individual 
exposures were sky-subtracted, flatfielded, aligned, and coadded.

The central cluster of NGC 3603 has been chosen for the initial 
focus alignment and the on-going focus monitoring (PI: G.\ Schneider, 
A.\ Suchkov) of the new NICMOS instrument at the Hubble Space Telescope.
Because of its proximity to the central cluster, Sher 25
is included in all HST/NICMOS observations with camera 3 (NIC3) 
centered on the cluster. 
We retrieved 18 calibrated images in the He{\sc i}\,1.08\,$\mu$m
line observed between May 6 and July 28 1997 from the
HST archive. The total integration time was 608s. We rotated all images 
to a common orientation.  Using our own IDL procedures, the individual 
images were then cross-correlated with each other in order to determine 
the offsets. Then the images were Fourier-shifted (to allow for the
sub-pixel alignment) and coadded. 
Due to technical problems with NIC3, the images were slightly out of focus. 
The coadded image has a resolution of 0\farcs7, whereas the best individual 
exposure had a resolution of 0\farcs5.

\section{Dynamical Structure of Sher25's Nebula}

\begin{figure*}[htb]
\centerline{\psfig{figure=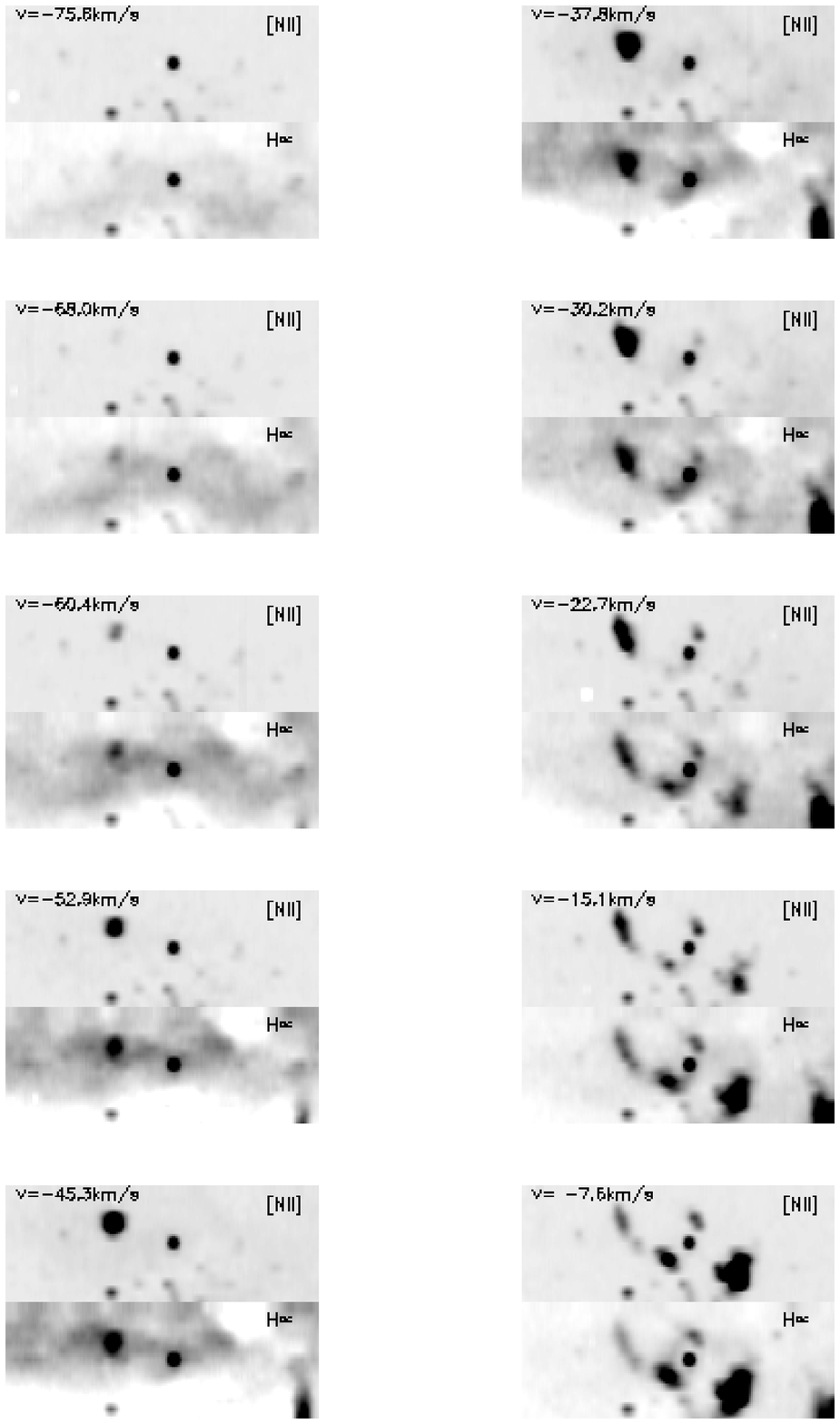,angle=0,width=14cm}}
\end{figure*}

\begin{figure*}[htb]
\centerline{\psfig{figure=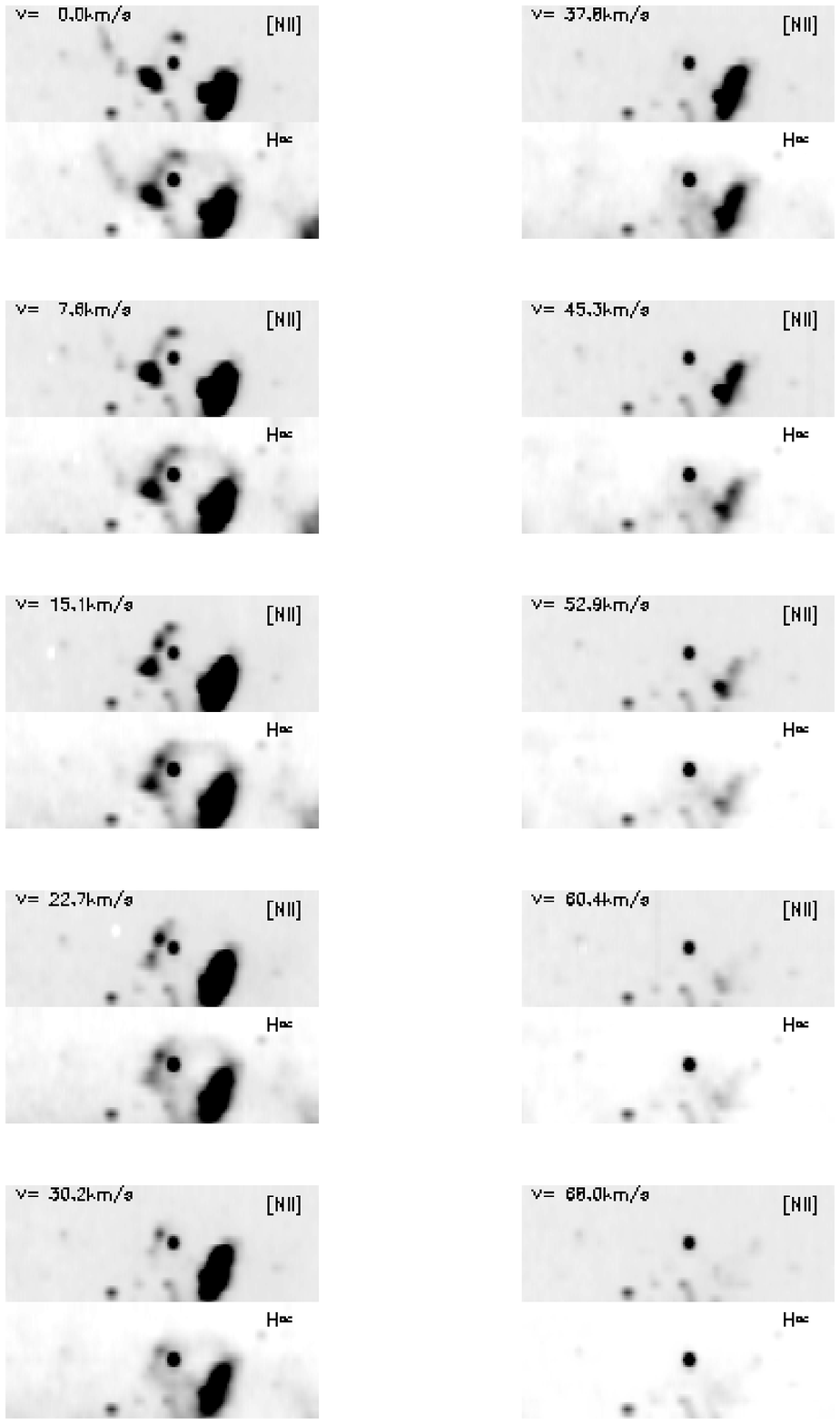,angle=0,width=14cm}}
\figcaption{Iso-velocity maps of Sher 25's circumstellar nebula in
[N{\sc ii}] and H$\alpha$. The field of view is $78'' \times 28''$, north
is up and east is to the left.
\label{fig1b}}
\end{figure*}

To study the morphology of the nebula in distinct velocity bins we
extracted iso-velocity maps from the 3-D data cube. The maps clearly 
trace the expanding ring and the bipolar lobes. In Figure 1 
(Plate x) we show iso-velocity maps in the [N{\sc ii}]\,658.4\,nm line
(upper panel) and H$\alpha$ line (lower panel).
The velocity offset between individual maps is 7.556 km\,s$^{-1}$. 
The zero point was chosen in such a way that the 
line of nodes of the ring has a velocity of 0 km\,s$^{-1}$.
A non-uniform background arising from the giant H{\sc ii} region NGC 3603
is present in almost all of the H$\alpha$ maps.

The overall structure of Sher 25's nebula has the shape of an
hourglass with the inner ring at the waist of the hourglass.
The ring is clumpy and coplanar. It is most prominent
in the velocity channels between $-30.2$ km\,s$^{-1}$ and $+30.2$ km\,s$^{-1}$.
The bright bipolar lobes lie on the surface of the hourglass.
Its silhouette can be traced on individual iso-velocity maps. The 
north-eastern silhouette is best visible in the velocity range of
$-22.7$ km\,s$^{-1}$ to $-7.6$ km\,s$^{-1}$, whereas the south-western 
silhouette is best visible in the velocity range of $+7.6$ km\,s$^{-1}$ to 
$+22.7$ km\,s$^{-1}$. Each of these bipolar lobes is pretty much complete.
In addition, a compact, high-velocity ``bullet'' is present in each lobe.
The bullet-type knots have velocities of up to $\pm 60.4$ km\,s$^{-1}$.
In some of the iso-velocity
maps a faint linear structure seems to connect the south-western
``bullet'' with Sher 25 (most prominent from $+7.6$ km\,s$^{-1}$ to $+30.2$
km\,s$^{-1}$).
Apart from these high-velocity components, the tips of the bipolar
lobes are restricted to projected velocities of $\pm 30$ km\,s$^{-1}$.

The two halves of the hourglass are asymmetric with respect to the
plane of the ring. While both lobes extend to equal
distances above and below the plane, their shapes are different.
The north-eastern lobe is prolate and the
south-western lobe is oblate with respect to the polar axis.
The bipolar caps show a rather complex structure. The south-western
polar cap is beret-shaped and is seen almost edge-on. The bright rims
of the cap can be seen easily on H$\alpha$ images (cf.\ Brandner et al.\
1997, Fig.\ 2/Plate L9). The beret shape might result from ram
pressure as the whole nebula appears to
be moving to the south-west relative to the interstellar medium (see below).

\begin{figure}[htb]
\centerline{\vbox{\psfig{figure=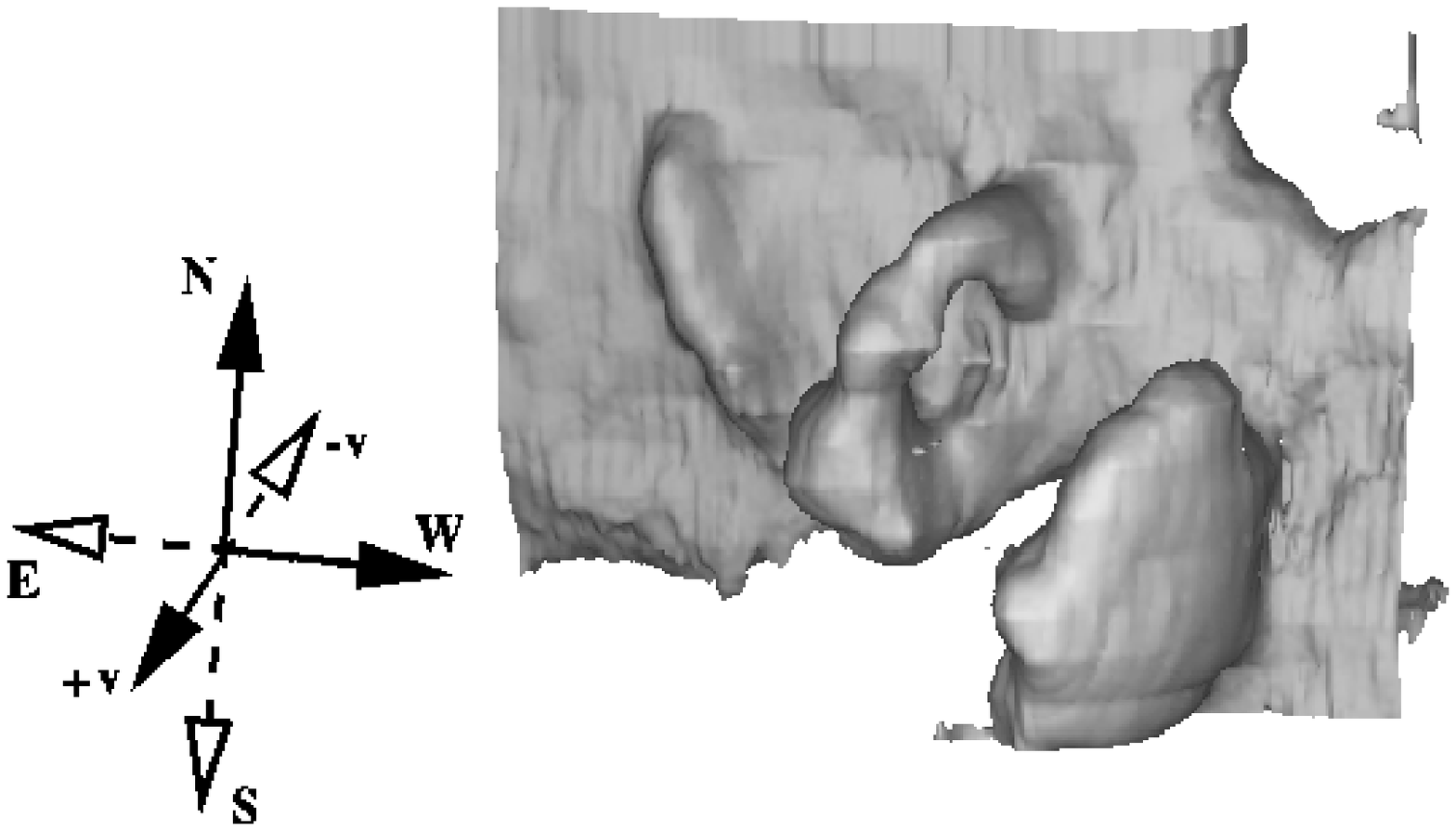,angle=270,width=8cm}
                  \psfig{figure=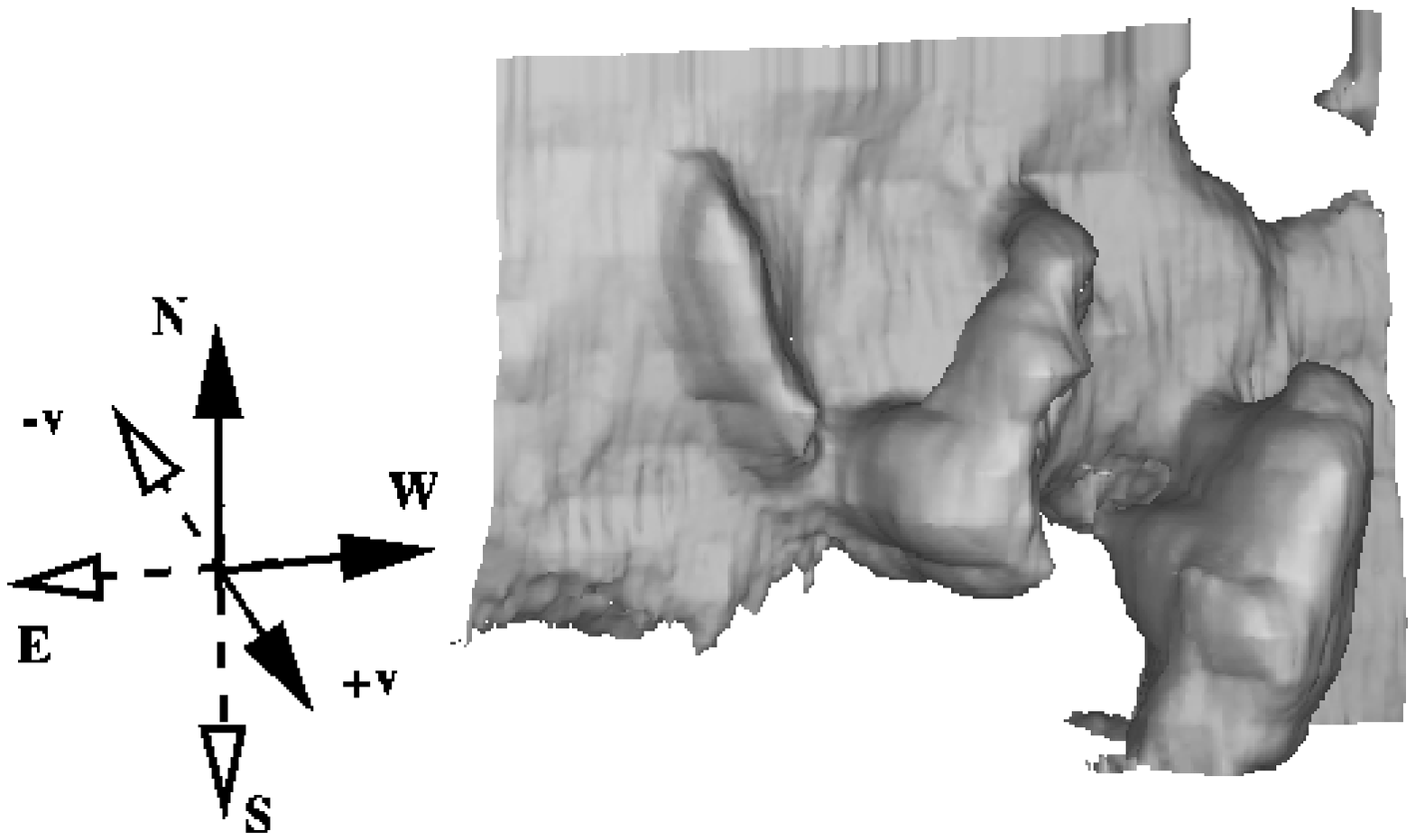,angle=270,width=8cm}}}
\figcaption{H$\alpha$ iso-surface plots of the 3-D data cube of the
nebula around Sher 25 from two different viewing angles (215$^\circ$, top \&
165$^\circ$, bottom). Note that we show the nebula as seen from `behind'
  to prevent the redshifted portions of the nebula (+v)
being obscured through the underlying blueshifted H{\sc ii} region.
The field of view is $52'' \times 28''$ ($1.6 {\rm pc} \times 0.8 {\rm pc}$)
and the velocity ranges from $-94$ km\,s$^{-1}$ to $+94$ km\,s$^{-1}$.
Bright stars have been subtracted.
\label{fig2}}
\end{figure}

Assuming that all nebular material has been ejected by Sher 25
instantaneously, the radial velocity of the material is directly
proportional to its distance
from Sher 25 along the line of sight. Tomographic imaging techniques can
then be applied to visualize the 3-D structure of the nebula. In Figure 2
we show the nebula as seen from different viewing angles (215$^\circ$ \&
165$^\circ$).
 
As can be seen from Figure 3, Sher 25 itself is displaced to the southwest
with respect to the center of the inner ring. This and the asymmetry in the
bipolar lobes indicate that Sher 25 and the whole outflow structure might be
moving relative to the ambient interstellar medium.
We confirm that the ring has an inclination angle of $\approx$ 65$^\circ$ 
as derived by Brandner et al.\ (1997). The expansion
velocities of the ring and the bipolar lobes, however, differ slightly
from previous estimates. The discovery of the expanding ring and
the velocity estimates were based on a single long-slit spectra
across the north-eastern lobe and the ring (Brandner et al.\
1997). Our new observations cover the entire hourglass nebula and
yield a deprojected expansion velocity of 30 km\,s$^{-1}$ for the ring. 
The bipolar lobes show deprojected velocities 
of 70 km\,s$^{-1}$, whereas the high-velocity knots are moving with 
velocities up to 140 km\,s$^{-1}$.

With these improved velocities, we derive a dynamical age of 6560 yr for the 
ring (diameter 0.4 pc), and a dynamical age of 6700 yr for the bipolar 
lobes, which extend to about 0.7 pc above and below the plane of the ring. 
Thus, both the ring and the bipolar lobes can be traced back to the 
same mass ejection event, which occurred approximately 6630 yr ago
(averaging the two ages).
The high-velocity knots might be younger ($\approx$ 3400 yr).

\section{Physical Properties of Sher 25's Nebula}

\begin{figure*}[htb]
\centerline{\psfig{figure=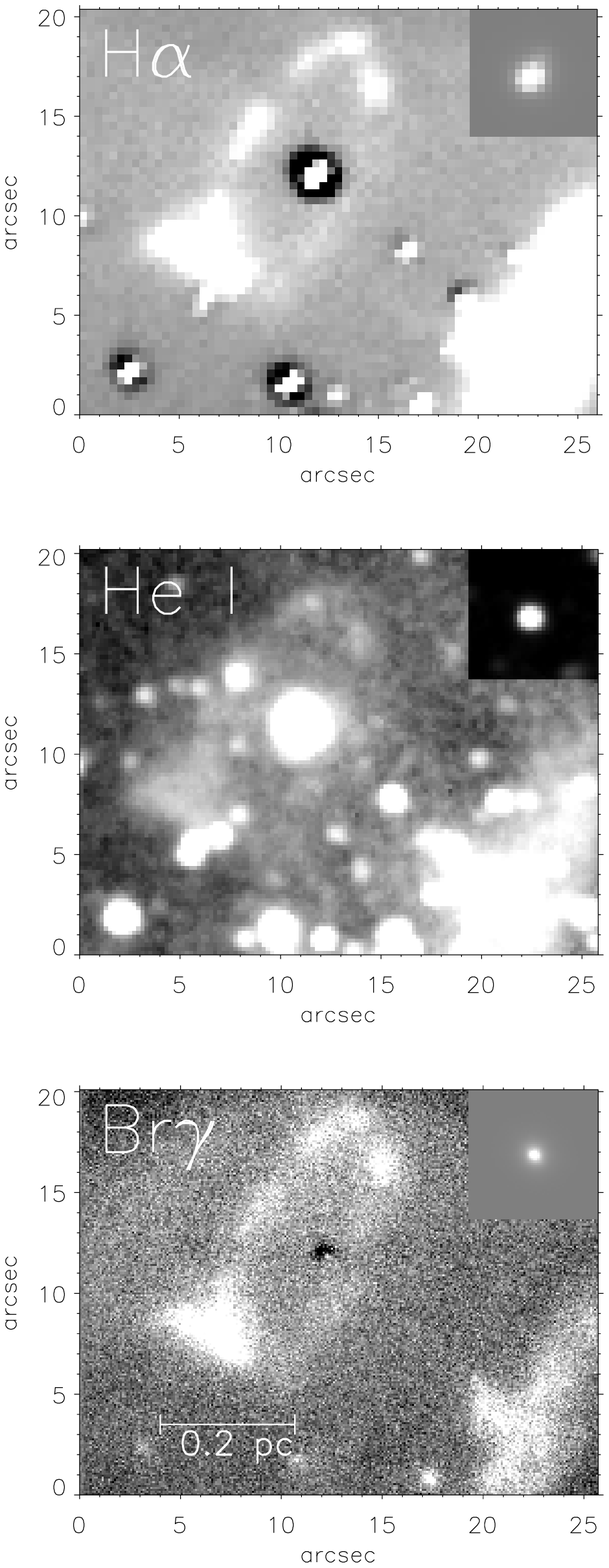,angle=0,width=13cm}}
\figcaption{Emission-line images of Sher 25's ring. Top: Continuum-subtracted 
H$\alpha$ image obtained with direct imaging (cf.\ Brandner et al.\ 1997). 
Middle: He{\sc i}\,1.08\,$\mu$m
observations (not continuum-subtracted) obtained with HST/NICMOS (NIC3).
Bottom: Continuum-subtracted Br$\gamma$ image obtained with adaptive optics.
The insert in the upper right corner of each panel shows the point spread 
function. Note that the width of the ring is clearly resolved
on our near-infrared adaptive optics and HST images. A faint linear structure
is visible between Sher 25 and the south-western cap.
North is up and east is to the left.
\label{fig3}}
\end{figure*}

Figure 3 (Plate xx) shows emission line images of Sher 25's ring in the
H$\alpha$ (top), He{\sc i}\,1.08\,$\mu$m (middle), and  Br$\gamma$
(bottom) lines. 
The width of the ring, 0\farcs9 (0.027 pc or $8\times 
10^{16}$ cm, i.e., $\delta$R/R=1:8), 
is clearly resolved in the near-infrared adaptive optics 
images in the Br$\gamma$ line and the HST/NICMOS images in the 
He{\sc i} line. 

No temperature-sensitive line combinations were available in our spectra.
We thus assumed a temperature of 10$^4$ K which is typical for 
H{\sc ii} regions. Electron densities were computed in two independent
ways, based on the H$\alpha$ surface brightness and using the 
[S{\sc ii}]\,671.7\,nm\,/\,[S{\sc ii}]\,673.1\,nm line ratios. Both calculations
agreed within a factor of two.
The electron density in the ring is around 10$^3$
cm$^{-3}$ in the bright knots and drops to  500 cm$^{-3}$ in the faint 
filaments between bright knots. Only the bright feature at the southern
rim of the ring shows a higher electron density of
$1.8 \times 10^3 {\rm cm}^{-3}$.
In the brightest knot of the north-eastern lobe
we measure an average electron density of around 
$1.5 \times 10^3 {\rm cm}^{-3}$, with a peak value of 
$3 \times 10^3 {\rm cm}^{-3}$. The south-western 
lobe has a mean electron density of $1.6 \times 10^3 {\rm cm}^{-3}$.

The masses deduced from the free electron densities depend strongly
on the geometry of the nebular material.
The adaptive optics image allows us to resolve the width of the ring. 
The beret-shaped polar cap at 
the top of the south-western lobe is resolved into two layers that
correspond to the receding and approaching rim of the cap, respectively.
The masses given in Table 1 assume a depth comparable to the
minor axis (lower limit) or a depth comparable to the major axis 
(upper limit) for the knots. The total ionized mass in the ring and the
lobes is $\le$ 0.7 M$_\odot$. Since this is small compared to the total
mass loss of a massive star over its lifetime, and the dynamical age of the
ring and the lobes are similar, we conclude that the mass ejection
event occurred during a brief and violent phase with a very 
high mass-loss rate.

\begin{deluxetable}{lrccc}
\tablecaption{Electron density, mass, and velocity in the ring
and outflows around Sher25.\label{tbl-1}}
\footnotesize
\tablehead{\colhead{Region}
&\colhead{N$_{\rm e}$ [cm$^{-3}$]}    & \colhead{M/M$_\odot$} & \colhead{v [km\,s$^{-1}$]} }
\startdata
 Inner ring & 500--1800   & 0.01-0.10  & 30 \\
 NE lobe & 1500-3000   & 0.02-0.07 & 70 (140)$^a$ \\
 SW lobe & 1600-3000   & 0.25-0.50 & 70 (140)$^a$ \\
\enddata
$^a$expansions velocity of the bullets
\end{deluxetable}

The smallest [N{\sc ii}]/H$\alpha$ ratio can be found in the ring
where it varies between 0.5:1 and 0.7:1. Only the bright knot at the southern
rim of the ring has a higher ratio of 1.1:1. The [N{\sc ii}]/H$\alpha$ 
ratio is considerably larger in the bipolar lobes, with a peak value of
1.7:1 in the north-eastern lobe and around 2:1 in the south-western lobe.
Thus, regions with the highest electron density 
also exhibit the largest [N{\sc ii}]/H$\alpha$ ratio.

\section{Two of a Kind? -- SN\,1987\,A and Sher 25}

How does Sher 25's circumstellar nebula compare to the
inner and outer rings around SN\,1987\,A? 

{\bf Ionization source:} The progenitor of SN\,1987\,A
had a Morgan-Keenan (MK) type of B3\,I (e.g., Walborn et al.\ 1989) and 
thus did not provide enough ionizing flux
for the inner and outer rings to become visible before the flash ionization
by the supernova explosion. Sher 25 has an MK type of B1.5\,I (Moffat 1983). 
In addition, Sher 25 is located just 20$''$ (0.6 pc projected
separation) north of HD 97950, the core of the young massive cluster at the 
center of NGC 3603. The fact that the brighter regions of the bipolar
lobes are facing the cluster indicates that ionizing photons
and/or fast stellar winds from the cluster's O and WR stars (e.g., Drissen
et al.\ 1995) provide an external ionization source. The cap at the tip
of the south-western lobe faces the direction of motion (see below).
Compression might be responsible for the higher flux observed here.

{\bf Morphology:} High spatial resolution observations of SN\,1987\,A
(e.g., Wampler et al.\ 1990; Plait et al.\ 1995; Panagia et al.\ 1996) revealed
the presence of an inner ring with a diameter of 0.4 pc and two outer rings 
0.4 pc above and below the plane of the inner ring (Cumming \& Lundqvist 1997).
 The width of the inner
ring appears to be $\le$ 0\farcs12 (0.030 pc at 50 kpc; Plait et al.\ 1995) and 
the outer rings are 
unresolved down to 0\farcs05, (0.013 pc at 50 kpc; Panagia et al.\ 1996).
While the outer rings might sit on the top and bottom of an hourglass-shaped
nebula, with the inner ring around its waist, no ionized material has 
been detected in the interior of the hourglass. For Sher 25, the ring has a 
diameter of 0.4 pc and a width of 0.027 pc, and the polar lobes 
stretch out to 0.7 pc above and below the central plane. The complete
silhouette of the hourglass can be seen in emission in the H$\alpha$ and
the [N{\sc ii}] lines.

{\bf Free electron density and mass:}
In the case of SN\,1987\,A, which lacks any external or
internal ionization source besides the supernova explosion itself,
only the inner ring and the two outer rings are ionized and showed
peak electron densities up to 10$^4$ cm$^{-3}$ shortly after the light
from the supernova reached the rings (e.g., Fransson et al.\ 1989; Plait et
al.\ 1995). The majority of the material within the hourglass is neutral.
The mass in each ring is less than 0.008 M$_\odot$
and the total mass in the hourglass amounts to 0.34 M$_\odot$ to 1.7
M$_\odot$ (e.g., Crotts \& Kunkel 1991; Burrows et al.\ 1995).
The total ionized mass within Sher 25's nebula is of the order
of 0.3 M$_\odot$ to 0.7 M$_\odot$.

{\bf Kinematics:} 
The expansion velocities of the inner ring around SN\,1987\,A and
Sher 25 are 10 km\,s$^{-1}$ and 30 km\,s$^{-1}$, respectively. The bipolar 
lobes recede from Sher 25 with about 70 km\,s$^{-1}$ whereas the outer shell
around SN\,1987\,A expands with a velocity of 85 km\,s$^{-1}$
(Burrows et al.\ 1995). Based on dynamical time scales, the inner ring 
around SN\,1987\,A is about 10$^4$ yr younger than the outer rings, and it 
also shows a stronger enrichment in heavy elements (Panagia et al.\ 1996).
The bipolar lobes and the ring 
around Sher 25 appear to be coeval.

{\bf [N{\sc ii}]/H$\alpha$ ratio:} For SN\,1987\,A the highest
[N{\sc ii}]/H$\alpha$ ratios were observed in the inner ring, up to 4.2:1,
and smaller ratios in the outer rings, 2.5:1 (Panagia et al.\ 1996). 
Sher 25 shows an opposite
behavior, the line ratios being higher in the bipolar outflows, up to 2.1:1,
than in the central ring, 1:1. Thus while the ring around Sher 25 
has similar brightness in H$\alpha$  and [N{\sc ii}], the inner ring around
SN 1987A is 1.5$^{\rm m}$ brighter in [N{\sc ii}] than in H$\alpha$.

\section{Conclusion}
 
Do the nebulae around Sher 25 and SN\,1987\,A belong to the same 
class of objects?
There are many morphological similarities between the bipolar nebulae around
SN\,1987\,A and Sher 25. 
We suggest that they indeed represent the first members of a new
class of bipolar nebulae around blue supergiants in their final evolutionary
stages. They might be situated between Luminous Blue Variables and 
Planetary Nebulae.

We will learn more about the structure and abundance of SN\,1987\,A's
inner ring once the supernova remnant impinges upon the inner ring
in the year 2005$\pm$3 (Chevalier \& Dwarkadas 1995; Pun \& Kirshner 1996). 
In the meantime, detailed follow-up studies and
theoretical modeling of the ring and hourglass nebula around Sher 25
should greatly advance our knowledge of ring structures and bipolar nebulae
around blue supergiants. Compared to the rings around SN\,1987\,A,
Sher 25 offers the advantage that it can be studied
at 8 times higher spatial resolution.
The hourglass structure is continuously ionized and will not
rapidly fade away or be destroyed like SN\,1987\,A's rings. Forthcoming
high spatial resolution observations of Sher 25 using the HST/WFPC2
(in Cycle 7) will enable us to study the ring structure
in more detail. Understanding
the formation of the nebula and analyzing its abundances will also shed light
on the surface enrichment and mass loss history of the central star.

\acknowledgements
YHC acknowledges the NASA grant NAG 5-3246. EKG acknowledges 
support by the German Space Agency (DARA) under grant 05 OR 9103 0.

\end{document}